\documentclass[]{spie}  

\newcommand{\rhos}{\rho_\odot}
\newcommand{\rhop}{\rho_\oplus}
 
\usepackage{amsmath,amsfonts,amssymb}
\usepackage{graphicx}
\usepackage[colorlinks=true, allcolors=blue]{hyperref}

\title{RISTRETTO: a VLT XAO design to reach Proxima Cen b in the visible}

\authorinfo{Further author information, send correspondence to N. Blind (nicolas.blind@unige.ch)}

\author[a]{N. Blind}
\author[a]{M. Shinde}
\author[a]{I. Dinis}
\author[a]{N. Restori}
\author[a]{B. Chazelas}
\author[b]{T. Fusco}
\author[c,d,e]{O. Guyon}
\author[f]{J. K\"uhn}
\author[a]{C. Lovis}
\author[g]{P. Martinez}
\author[h,b]{M. Motte}
\author[b]{J.-F. Sauvage}
\author[g]{A. Spang}

\affil[a]{D\'epartement d'astronomie, Universit\'e de Gen\`eve, Versoix, Switzerland}
\affil[b]{DOTA, ONERA, 13661 Salon cedex AIR, France}
\affil[c]{University of Arizona, Steward Observatory, Tucson, Arizona, United States}
\affil[d]{National Astronomical Observatory of Japan, Subaru Telescope, National Institutes of Natural Sciences, Hilo, HI96720, USA}
\affil[e]{Astrobiology Center, National Institutes of Natural Sciences, Osawa, Mitaka, Tokyo, JAPAN}
\affil[f]{Space Sciences Institute, University of Bern, Bern, Switzerland}
\affil[g]{Observatoire de la C\^ote d’Azur, CNRS, Laboratoire Lagrange, Nice, France}
\affil[h]{Aix Marseille University, CNRS, CNES, LAM, Marseille, France}

\begin{document}
\maketitle

\begin{abstract}
RISTRETTO is the evolution of the original idea of coupling the VLT instruments SPHERE and ESPRESSO \cite{lovis_2016a}, aiming at High Dispersion Coronagraphy. RISTRETTO is a visitor instrument that should enable the characterization of the atmospheres of nearby exoplanets in reflected light, by using the technique of high-contrast, high-resolution spectroscopy. Its goal is to observe Prox Cen b and other planets placed at about 35mas from their star, i.e. $2\lambda/D$ at $\lambda$=750nm. The instrument is composed of an extreme adaptive optics, a coronagraphic Integral Field Unit, and a diffraction-limited spectrograph (R=140.000, $\lambda =$620-840 nm).

We present the RISTRETTO XAO architecture that reach the specification, providing contrasts down to $5\times10^{-5}$ at 2$\lambda/D$ from the star in the visible, in the presence of atmosphere and low wind effect. This performance is allowed by a new two-sensors-one-dm architecture, some variations to the already known concepts of unmodulated pyWFS and zWFS, and exploiting to the maximum of their capabilities the state-of-the-art high speed, low noise cameras \& fast DM. We present the result of end-to-end simulations, that demonstrate stable closed loop operation of an unmodulated pyramid and a zernike WFS (together), and in presence of low wind effect. \end{abstract}


\section{INTRODUCTION}
\label{sec:intro}  

Direct detection of the light from extrasolar planets is a challenging objective. For young exoplanets on wide orbits this is achieved by high-contrast imaging and coronagraphy. For close-in exoplanets it is possible to take advantage of planetary transits and obtain a direct measurement of the IR flux of giant planets using secondary eclipses and phase curves. However, many known exoplanets including those orbiting the brightest and nearest stars remain out of reach of these techniques.

The RISTRETTO instrument will attempt reflected-light observations of spatially-resolved exoplanets for the first time. The proposed technique for RISTRETTO is High Dispersion Coronagraphy (HDC) spectroscopy, that can enable the $10^{-7}$ contrast level. To separate the planet from the star, adaptive optics on an 8m class telescope is used. In order to reach the required contrast, a coronagraph and an extreme AO stage are needed. Using a high-resolution spectrograph, the radial velocity shift between star and planet allows us to separate the stellar and planetary spectral lines, providing an additional factor $\sim$1000 in achievable contrast. Other experiments are being developed based on a similar approach. However they are targeting the thermal emission of young, massive planets at larger separations in the IR. Driven by the Proxima b science case, the RISTRETTO instrument will be targeting reflected light in the visible wavelength range. The short wavelengths allow us to spatially resolve Proxima b and other known very nearby exoplanets, placing them at about 2 $\lambda/D$ on a 8m-class telescope.

We focus in this paper on the XAO concept, validated by advanced end-to-end simulations, with support of our experience and lab characterisation. The coronagraphic IFU, based on a new PIAA-Nuller concept, was introduced in \cite{blind_2022a} and is now being tested in the lab \cite{restori_2024a}.


\section{XAO requirements and error budget}
\label{sec:xao}

In the case of a photon-noise limited measurement, the SNR of HDC technique is given by \cite{lovis_2016a}:
\begin{equation}
    \mathrm{SNR} \varpropto \sqrt{T} \dfrac{\rhop}{\sqrt{\rhos}},
\end{equation}
where $T$ is the total transmission, and $\rhop$ and $\rhos$ are respectively the planet and the stellar light coupled into the external fibers of the IFU. Note that in the rest of the paper, we will not consider contrast anymore, since it mixes the planet and star coupling, which are in practice optimized almost individually.

\noindent The requirements for RISTRETTO are the following:
\begin{enumerate}
    \item $\rhop \ge 50\%$ for host stars as faint as PDS70 (Imag =11).
    \item $\rhos \le 10^{-4}$ for Proxima Cen
\end{enumerate}
, applying for median seeing conditions (Seeing = 0.83" at zenith distance of $30^\circ$, $L_0$ = 20m, Wind = 9.5m/s). Note that maximum elevation of Proxima Cen is 52deg, i.e. a corrected median seeing $\sim$0.90”.

The corresponding XAO requirements can summarize as follow:
\begin{itemize}
    \item Strehl($\lambda$=750nm) $\ge$ 70\% or WFE $\le$ 70 nm RMS;
    \item Low order WFE, within 3 cycles, $\le$ 10nm RMS, including pupil fragmentation effects;
\end{itemize}
This can be answered with the minimum XAO configurations from Tab.~\ref{tab:XAO_config}, although we aim at even more performing ones to allow higher performance in various situations and higher observing flexibility. 

\begin{figure}[b]
    \centering
    \begin{minipage}{0.43\textwidth}
    \centering
    \includegraphics[width=1.\textwidth, trim={0.5cm 0.3cm 1cm 0}, clip]{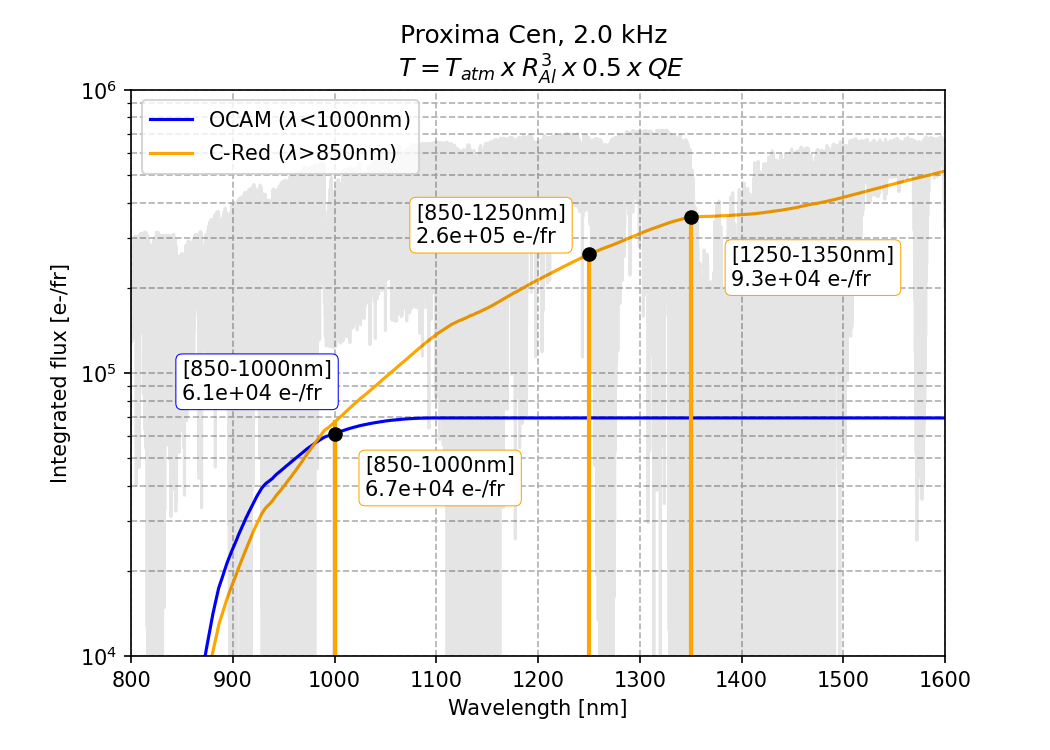}
    \caption{Photon budget on Prox Cen for 2kHz system.}
    \vspace{0.9cm}
    \label{fig:wfs_photon}
    \end{minipage}
    \hfill
    \begin{minipage}{0.54\textwidth}
    \centering
    \includegraphics[width=1.\textwidth]{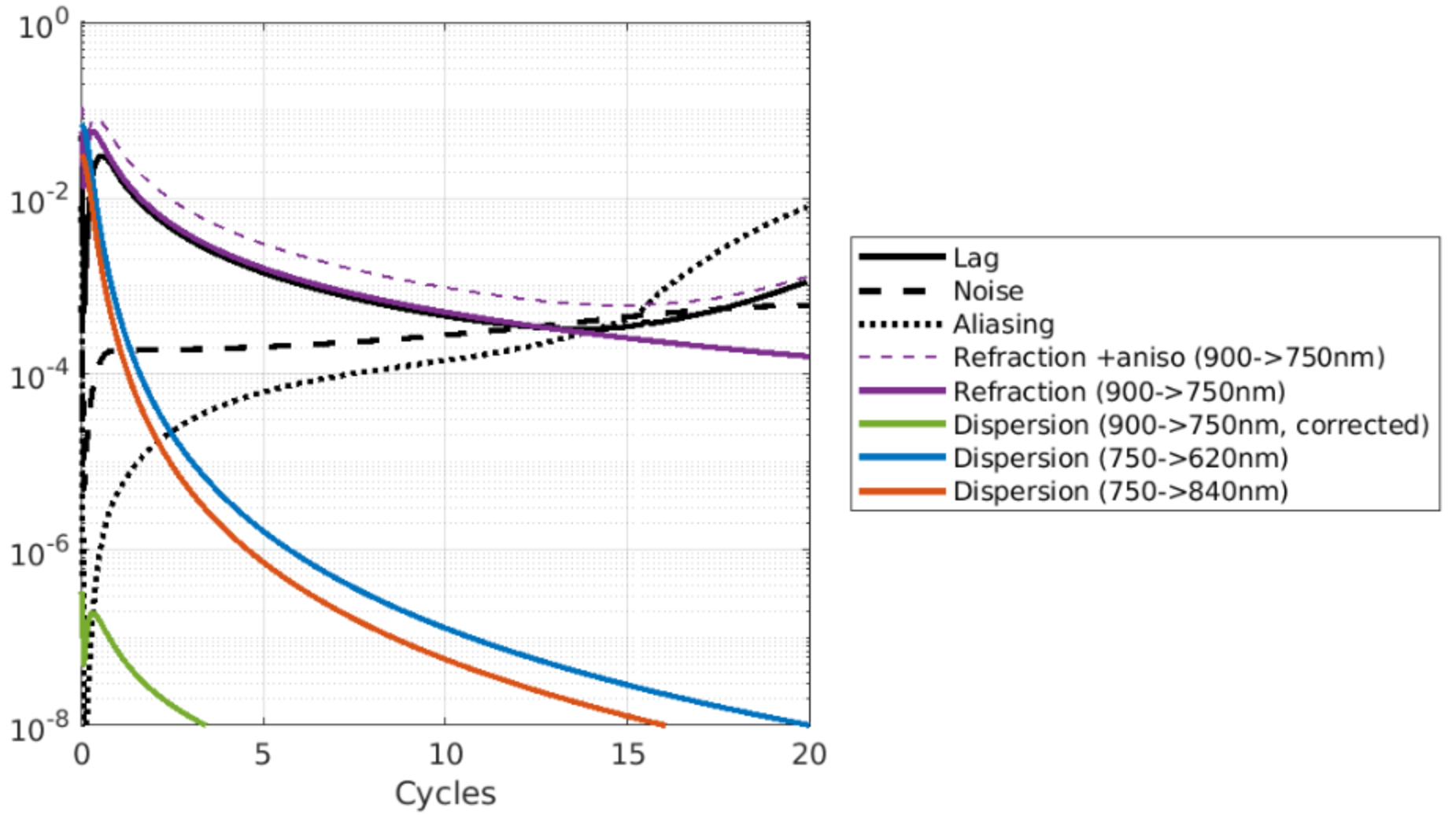}
    \caption{AO PSDs profiles for the different term in the AO error budget for a 50x50, 4kHz system under median seeing and airmass=2. Chromatic terms computed for $\lambda_{eff}^{WFS} = 900$ nm.}
    \label{fig:psd}
    \end{minipage}
\end{figure}

While a modulated pyWFS can potentially reach the required performance regarding sensitivity/photon noise considering a broadband 850-1250nm (Fig.~\ref{fig:wfs_photon} \& Tab.~\ref{tab:flux_WFS}), it offers little margin in case of bad conditions as well as basically no sensitivity to pupil fragmentation effects. Extending the band further to the red would increase the impact of chromatic terms. An unmodulated pyWFS (or Zernike WFS) on the other hand, being about 10 times more sensitive on low orders, offers about 2 magnitude margin for any bandwidth $\Delta\lambda>100$nm in NIR, while offering potential to control fragmentation effects.

\begin{table}[b]
    \centering
    \caption{RISTRETTO XAO configurations following Fourier analysis}
    \vspace{0.3cm}
    \label{tab:XAO_config}
    \begin{tabular}{l|cccl}
        Configuration & $f_{loop}$ [kHz] & Delay [frame] & Nact & WFS \\
        \hline
         Minimum   & 1.5 & 2 & 32       & Mod. pyWFS \\
         Baseline  & 2.0 & 2 & $\ge$ 40 & Unmod. pyWFS \\
         Goal      & 4.0 & 1 & $\ge$ 50 & Unmod. pyWFS \\
    \end{tabular}
\end{table}

In the case of unmodulated pyramid, servo-lag is largely dominating the error budget at low spatial frequencies $\le 3\lambda/D$. For modulation of 2 to 3 $\lambda$/D, WFS noise on Prox Cen is on par with the lag term at low spatial frequencies.

The goal design will provide a significant boost in performance, potentially allowing to reach specification down to airmass=2 in median seeing conditions (i.e. for equivalent seeing $\sim$ 1.2"). This is important since Prox Cen airmass $\ge$ 1.25 in Paranal. A more ambitious configuration will not improve much performance, as we reach the limits of the coronagraphic IFU itself\cite{blind_2022a}.

Another limit of the baseline and goal configurations will ultimately be differential refraction between WFSing and science bands, generating chromatic pupil shift. At maximum Prox. Cen elevation (52$^\circ$), WFS effective wavelength $\lambda_{eff}^{WFS}$ should not exceed 1100nm to be at the level of the lag term. Under median seeing conditions, goal configuration can potentially reach specification at an airmass of 2 if $\lambda_{eff}^{WFS} \le 900$nm (Fig.~\ref{fig:psd}).

\begin{table}[h]
    \centering
    \caption{Minimum flux level on a pyramid WFS to ensure WFE $\le$ 10nm RMS within 3 cycles, for different modulation amplitudes. Readout noise is negligible.}
    \vspace{0.3cm}
    \label{tab:flux_WFS}
    \begin{tabular}{ll|ccccc}
        Modulation radius & [$\lambda/D$] & 0 & 1 & 2 & 3 & 5\\
        \hline
        Minimum flux & [e-/fr] & 4$\cdot 10^3$ & 8$\cdot 10^3$ & 20$\cdot 10^3$ & 40$\cdot 10^3$ & 170$\cdot 10^3$ \\ \\
    \end{tabular}
\end{table}

\section{Wavefront sensing in the NIR}

Very early in the project, we identified the need for NIR WFS, so that a C-Red1 camera was acquired and recently characterized\cite{shinde_2024a}.

\subsection{A dual WFS architecture}

A modulated pyramid is now a proven solution when it comes to obtaining high Strehl and high contrast far from the star. It is a less obvious option for RISTRETTO, its narrow working angle requiring $f\ge2$kHz in the NIR, where modulator stroke could be too limited\cite{lozi_2022a}. This will also ultimately limit the potential offered by current state-of-the-art DMs, cameras and WFSs. Not being sensitive to fragmentation effects, it must also be supported by a 2$^{nd}$ dedicated WFS. Alternative solutions like the flip-flop technique\cite{engler_2022a} or filtered pyWFS \cite{levraud_2022a} might limit contrast performance in our case. A unique unmodulated pyramid at $\lambda_{eff}^{WFS}$ $\le$ 1200nm could potentially work, although fragmentation control to nm levels remains to be demonstrated with this sensor.

\vspace{0.3cm}
Our baseline relies on a 2-color WFS solution, where one or two of the sensors should be able to sense fragmentation:
\begin{itemize}
    \item \textbf{A Red WFS} (R-WFS; $\lambda \ge 1400$nm) is responsible of closing the loop and maintaining high Strehl, with frequency $\ge$ 1kHz. The best trade-off regarding sensitivity, dynamic range and loop stability is an unmodulated pyWFS. Going as red as possible comes with important benefits: \textbf{\textit{1)}} It increases the OPD dynamic range by a factor 2-3 with respect to visible sensors, allowing to close the loop under seeing higher than 1.5", as demonstrated by our simulations. \textbf{\textit{2)}} It strongly reduces non-linearities: in median conditions, optical gains reach values above 0.8 with a flat behavior over spatial frequencies\cite{shinde_2024b}. In operation, this allows a simple scalar gain correction, reducing the need for a gain scheduling camera\cite{chambouleyron_2021a}. High optical gains (which can be seen as a transmission factor) also makes a much better use of available photons. Simulations performed at shorter wavelengths also show higher residuals once the loop is closed (despite an optical gain correction), which we attribute to higher non-linearities and modal confusion.
    
    Since it is unmodulated, the quality of the pyramid roof and tip is critical, so that we are favoring a 3-sided pyramid solution (to prototype by 2025). This solution will also require slightly less pixel, to push a bit the C-Red 1 max fps.
    \item \textbf{A Blue-WFS} (B-WFS; $\lambda$=850-1000nm) is responsible of pushing contrast, by sensing closer from the science band, at frame rate $\ge$ 2kHz. zWFS is currently our main candidate, as it proves more efficient to sense discontinuities. More details are given in Sect.~\ref{sec:pupil_fragmentation}. 
\end{itemize}
A careful choice of the R-WFS bandpath will better exploit the chromatic information between the two sensors. It could help detect 2$\pi$-jumps in piston modes, or disentangle mechanical vibrations from atmospheric jitter.

We aim at putting the two sensors on the same C-Red 1 detector. Thanks to the limited number of pixels required, the C-red1 can then run at up to 5kHz in CDS mode. A preliminary Zemax design shows the optical feasibility of this idea (Sect.~\ref{sec:fe}). As a backup, the B-WFS could still use an independent EMCCD camera.

\begin{table}[t]
    \centering
    \caption{Relation between C-Red 1 configuration and required modulator requirement}
    \vspace{0.3cm}
    \label{tab:C-Red 1_speed}
    \begin{tabular}{lcccc}
        Window(s) size & Max. FPS & Used FPS & Duty cycle & Modulator \\
        \ [pixels] & [kHz] & [kHz] & & [kHz] \\
        \hline
        256x320 (full frame) & 1.75 & 1.75 & 0.50 & 3.50 \\
        114x114 + 60x60      & 4.80 & 2.00 & 0.78 & 2.55 \\
        \end{tabular}
\end{table}

\subsection{Some implications of the C-Red 1 duty cycle}

As pointed out in \cite{lozi_2022a} , the time required to read the C-Red 1 in CDS induces a dead integration time. This dead time (a.k.a. duty cycle) amounts to 50\% when the camera is used at its maximum frame rate, meaning that half the photons are lost. This is problematic for Prox Cen with a modulated pyWFS, where we already have little margin. This duty cycle can be accounted as a transmission factor, which, for a given fps, increases with max fps.

For a modulated pyramid, the duty cycle also involves that the modulator has to be over-designed to perform a full rotation during the effective exposure time of the camera. For a full frame C-Red 1 at maximum CDS frequency of 1.75kHz, the modulation must then happen at 3.50kHz (Tab.~\ref{tab:C-Red 1_speed}). Using a sub-window reduces this effect by reducing read-out time / increasing max fps. For RISTRETTO, sensing in the NIR,  the modulation radius could also be limited at 2kHz.

Another limitation comes in the context of the dual WFS we envision. If we want to make use of a unique C-Red 1, a modulator will likely limit the maximum XAO frequency to no more than 2kHz. To relax this constraint on the modulator, we may imagine to average frames acquired at 2kHz to get an 'effective' 1kHz frame rate: however the 2 dead acquisition times during the modulation is a source of strong WFE noise estimation, well above the requirements for RISTRETTO. Such an option is discarded.

\subsection{Low Wind Effect sensing}
\label{sec:pupil_fragmentation}
\subsubsection{A variation around the zWFS}

Low Wind Effect is due to local radiative cooling of the telescope spiders, leading to a phase discontinuity around the spiders. This effect usually takes the form of a series of segmented Piston/Tip/Tilt (PTT) modes over each of the 4 \textit{segments} of the pupil. Whether these modes are physical or the result of the control loop is still under debate \cite{pourre_2022a}. When we introduce a local spider cooling effect in our simulations, a Shack-Hartmann WFS is very prone to introduce such PTT modes after a couple of iterations, as already noted in \cite{pourre_2022a}. On the other hand, an unmodulated pyWFS only generates segment piston jumps when closing the loop, and only on rare occasions (which could be easily detected with our dual-WFS). We did not perform that test with a modulated pyWFS.

Due to its very narrow working angle, RISTRETTO must control those modes to better than 10nm PTP (!). RISTRETTO must be ready to manage such effect whatever its origin and spatial distribution. The critical part is the discontinuity itself, which a standard DM cannot perfectly fit. The rest of the disturbance is, to 1$^{st}$ order, a sum of segmented PTT modes (or any combination of DM modes) that the DM can handle.

Although sensitive to discontinuities in the pupil, unmodulated WFS suffers from am important crosstalk between left/right modes around a spider: in our simulations, it performs poorly at controlling them, even in absence of atmosphere and modal noise. Tests performed by injecting as disturbance, not a PTT mode, but its best fit by the DM (hence much smoother and allowing perfect correction) shows performance equivalent to simulations without the effect: this points towards a misinterpretation of the DM fit error by the unmodulated WFS around the spider. On the other hand, the zWFS appears perfectly capable to correct such modes at the required level (see Sect.~\ref{sec:perf}).

\begin{figure}[t]
    \centering
    \includegraphics[width=\textwidth, trim={0 0 0 0}, clip]{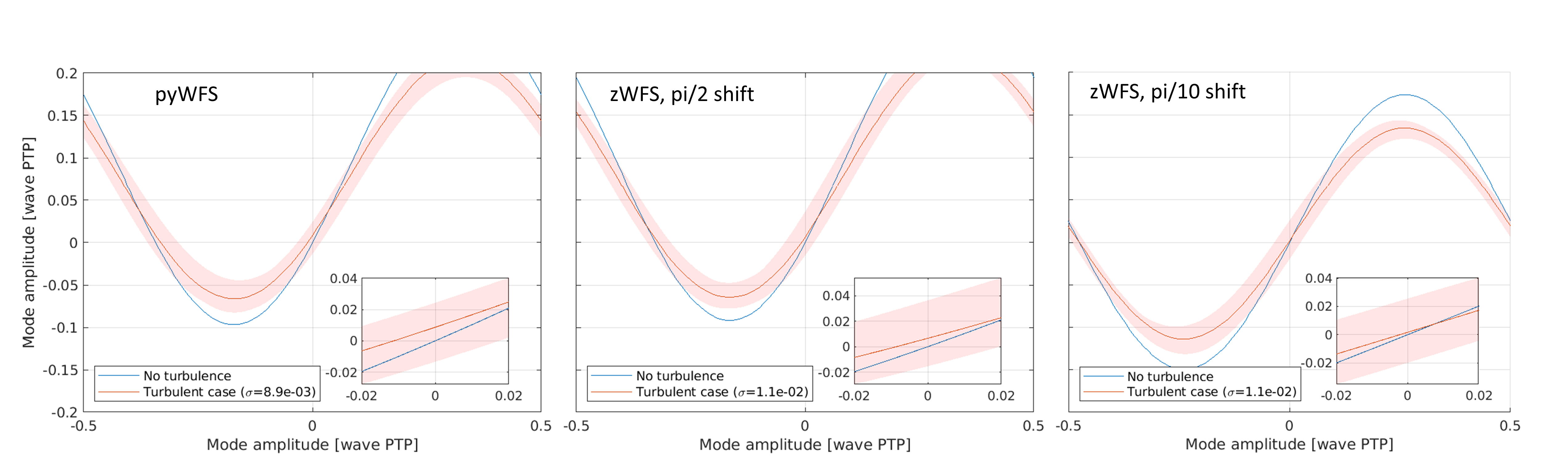}
    \caption{Petal response for pyWFS (left), $\pi/2$ zWFS (middle) and $\pi/10$ zWFS (right). Blue curve are natural response of the sensor to a pupil segment. The red curve (plus fill area) is the response in presence of residual turbulence in median conditions for our baseline design.}
    \label{fig:zwfs_vs_pyWFS}
\end{figure}

\begin{figure}[b]
    \centering
    \includegraphics[width=\textwidth, trim={0 4cm 12cm 0}, clip]{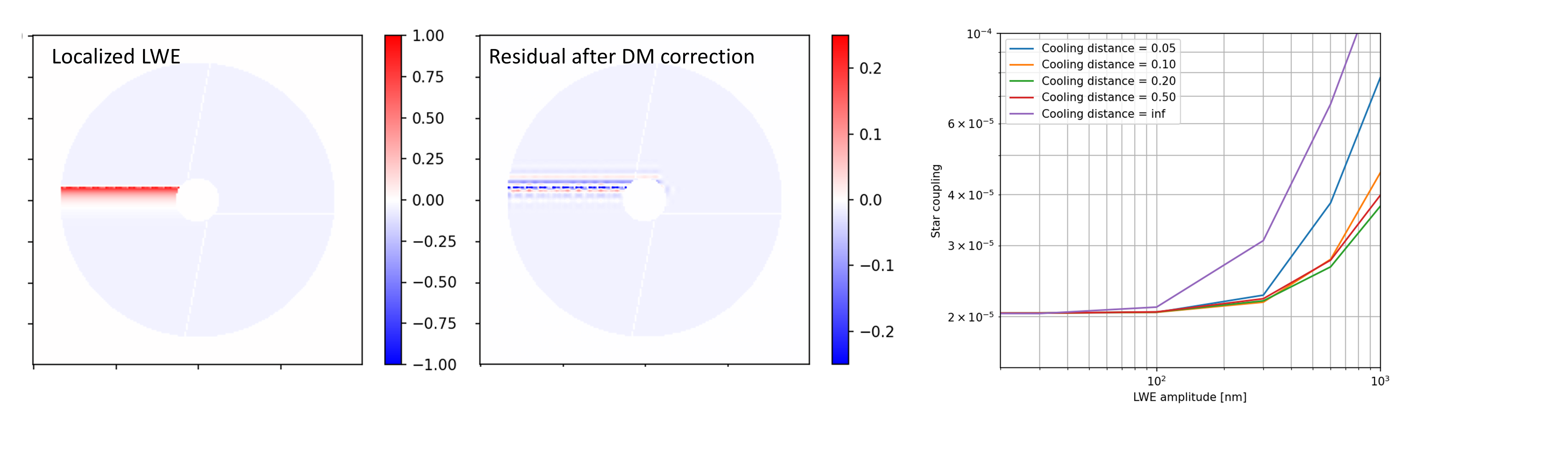}
    \caption{Impact of LWE on final performance. \textbf{Left}: localized LWE discontinuity. \textbf{Middle}: best residual after DM correction. \textbf{Right}: Impact of the best residual on PIAAN performance, as a function of the injected LWE amplitude, for different cooling zone size ('inf' corresponds to a full piston segment, therefore including 2 discontinuities).}
    \label{fig:lwe}
\end{figure}

 The standard zWFS with a $\pi/2$ phase offset suffers from an asymmetric dynamic range, which can be mitigated with a dual zWFS concept\cite{doelman_2019a, cisse_2023a}. It is not guaranteed yet that we can afford such a sensor due to limited space on the C-Red 1 and/or optical complexity or our will to push the C-Red 1 frame rate as high as possible. By reducing the phase to $\le\pi/6$, the response of the sensor tends to balance as well as to become less biased with turbulence or NCPA (Fig.~\ref{fig:zwfs_vs_pyWFS}), without impacting dynamic range, at the cost of sensitivity.

\subsubsection{The limit in LWE correction}

Ultimately, RISTRETTO performance is limited by the DM fitting error to local discontinuity along the spider. The impact on the PIAAN performance (Fig.~\ref{fig:lwe}) shows little sensitivity to the dimension of the effect: the discontinuity is the most problematic on that regard. The low frequency part of the disturbance plays a minor role, since within the correction capabilities of a standard DM. Injecting a single discontinuity of about 1$\mu m$ in amplitude (more or less what we expect from a couple degrees difference between spider and air\cite{sauvage_2015a}), the best correction by a DM reduces the PIAAN contrast by a factor $\sim$4. Increasing the actuator density from 40 to 60 or choosing a particular angle between spiders and DM actuators grid does not have a significant impact on the result.

The amplitude and duration of low wind effect is potentially low with respect to a RISTRETTO exposure for Prox Cen, and only appears in very good conditions, so that its integrated impact could be limited as long as the AO remains stable (see Sect.~\ref{sec:perf}).

\section{The deformable mirrors}
\label{sec:dm}

\subsection{Candidates}

We compared 3 different DMs architectures (Tab.~\ref{tab:dms}). Two are based on BMC 2k tweeter DM that would be supported by an ALPAO high stroke LODM. Those solutions are in principle simple to control: the HODM being also the fastest, it is the only one directly controlled by the WFSs. The LODM is then controlled by a high speed offload loop, at a frequency up to 500Hz, to maintain the HODM stroke usage to the minimum while not getting too close from the LODM resonance frequency.

The solution based on a BMC 2k-3.5 DM is our baseline, as well as the highest TRL option at the moment.

\begin{table}[h]
    \footnotesize
    \centering
    \vspace{0.5cm}
    \caption{Comparison of the 3 DM architectures considered for RISTRETTO}
    \vspace{0.3cm}
    \label{tab:dms}
    \begin{tabular}{p{1.cm}c|p{1.cm}c|p{3.5cm}|p{5cm}}
    \hline
    HODM & Role & LODM & Role & Pros & Cons \\
    \hline
    \hline
    BMC 2k-3.5 \newline(7$\mu m$)   & XAO & ALPAO 97 ($40\mu m$)& Offload 
    & - DMs of same size\newline - Simple WT control\newline - Highest stroke 
    & - Dead act. (potentially mitigated)\\
    \hline
    BMC 2k-1.5 (3.0$\mu m$) & XAO & ALPAO 292 ($22\mu m$)& Offload 
    & - DMs of same size\newline - Simple WT control\newline - 100\% yield guaranteed
    & - Higher order, lower stroke LODM: requires 'fast' offload TTM\newline
    - Limited HODM stroke for bad seeing, high airmass\\
    \hline
    ALPAO 64x64 (10$\mu m$) & Strehl & BMC 12x12 ($11\mu m$)& Contrast 
    & - No dead act.\newline - Highest Strehl, lowest $t_{int}$
    & - Large HODM/small LODM\newline - WT control more complicated (spatial \& temporal split)\newline
    - Lowest stroke: requires 'fast' offload TTM\newline
    - Most expensive\\
    \hline
    \end{tabular}
    \vspace{0.5cm}
\end{table}

\subsection{Dead actuator management}

Following our experience with KALAO DM(s), we developed some concern related to dead actuators with MEMS technology. However, the LODM and a couple of neighbouring HODM actuators could be used to minimize the imprint of the dead actuator. 3 situations were considered: 1 dead actuator, a cluster of 2 dead actuators, and 2 dead actuators at some distance through the pupil (Fig.~\ref{fig:dead_act}). In absence of turbulence, the residual error in those 3 cases little degradation, especially at low spatial frequency. With turbulence, it becomes negligible, while the cost on the HODM stroke appears also limited.

\subsection{DM dynamics}
\label{sec:dm_dynamics}
Measuring the dynamical behavior of DMs is a complex task, especially with the super fast MEMS technology developed by BMC. Some specialized benches are looking into this by using super fast cameras \cite{strobele_2022a}. 
If one is not interested in a precise characterisation of the spatial behavior of the membrane, another possibility is to measure the variation of flux coupled into a single-mode fiber as the DM moves. A simple photodiode can then measure the DM dynamics at several MHz. 

\begin{figure}[t]
    \begin{minipage}{0.55\textwidth}
        \centering
        \includegraphics[width=1\textwidth, trim={0 3cm 56cm 0}, clip]{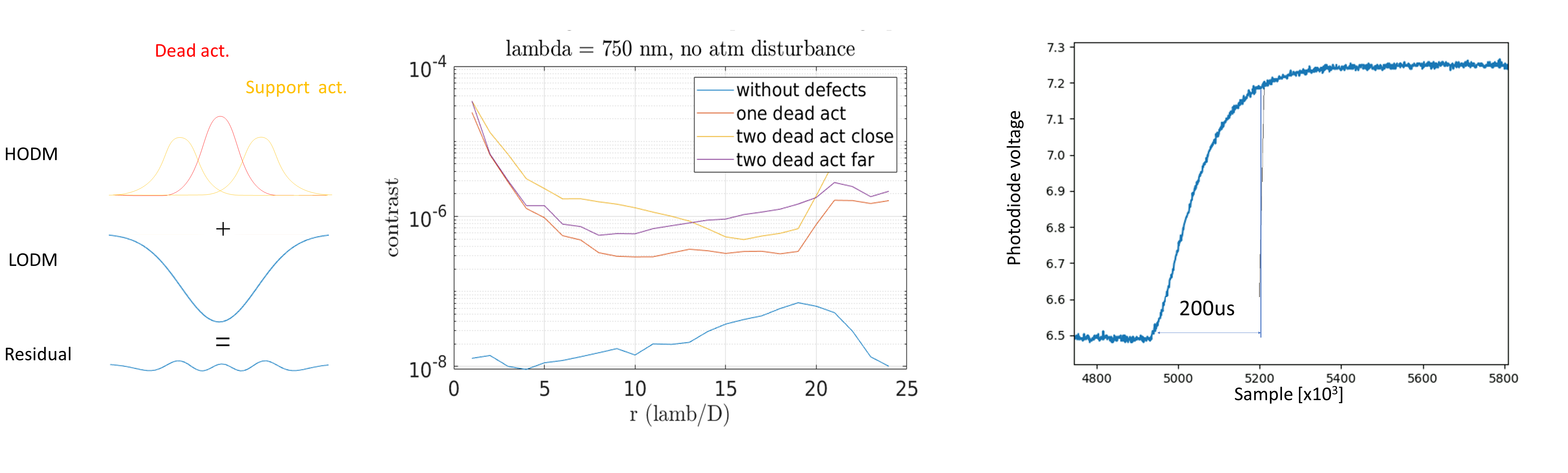}
        \caption{Dead actuator compensation strategy. \textbf{Left}: schematics of the compensation using the LODM as well as a couple of HODM actuators surrounding the dead one. \textbf{Right}: impact on imaging contrast for 3 different cases of dead actuators.}
        \label{fig:dead_act}
    \end{minipage}
    \hfill
    \begin{minipage}{0.40\textwidth}
        \centering
        \includegraphics[width=1\textwidth, trim={84cm 3cm 0cm 0}, clip]{Slide8.PNG}
        \caption{Dynamical characterisation of a Boston MultiDM using coupling into a singlemode fiber and a photodiode.}
        \label{fig:dm_dynamcis}
    \end{minipage}
\end{figure}

We performed such characterisation with the small BMC 12x12 MultiDM in our high contrast bench\cite{restori_2024a} (Fig.~\ref{fig:dm_dynamcis}). We could measure a 90\% rise time of $\sim 200 \mu s$ for KL or Zernike modes, slightly slower than the 150-170$\mu s$ from the data sheet. A finer analysis is probably required to remove a small parabolic trend of the coupling. 

Anyway, the absence of oscillation indicates a very stiff DM, with a resonant frequency of $\omega$=4.0kHz and a damping ratio $\zeta$=0.99. The 2k-3.5 DM presents the same characteristics on paper, so that a 4kHz control loop could be close from its actual limit.

\section{Control architecture}

\begin{figure}[b]
    \centering
    \vspace{0cm}
    \includegraphics[width=0.4\textwidth]{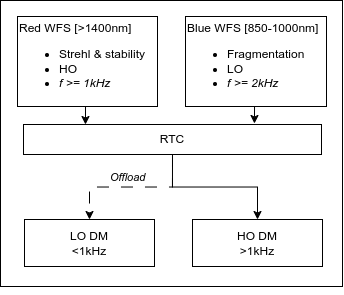}
    \caption{Schematic of the XAO architecture.}
    \label{fig:xao_architecture}
\end{figure}

At this stage of the project, we want to demonstrate the performance of the XAO with a simple integrator control loop. We allow the two WFSs to control the HODM (Fig.~\ref{fig:xao_architecture}) so that the control vector computes:

\begin{equation}
    c = M2C \times \left[  G_B \cdot RM_B \cdot s_B + G_R \cdot RM_R \cdot s_R \right]
\end{equation}

\noindent where subscripts $_B$ and $_R$ stand for the B-WFS and R-WFS. $s$ are the slopes/meta-intensities from both sensors, $RM$ the HODM KL reconstruction matrices from slopes of both sensors, and $G$ are diagonal matrices, containing modal loop gain and optical gain compensation of individual modes. $M2C$ is the modes to actuator control matrix.

We perform a spatial decoupling between the two sensors. The loop is first closed with the R-WFS only ($G_B=0$), and after $\sim$50-200ms, the B-WFS takes control of the 100 first low order modes so that $G_B$[1:100] $\sim 0.4-1.0$ and $G_R$[1:100] = 0. This must be tuned, but demonstrates stable control with a noticeable improvement in the center of the correction area over the R-WFS control alone. This is very preliminary work, and as reported by others, the zWFS made a poor work at controlling several hundreds of modes, loop diverging very quickly. Our chance could be our narrow working angle.

More advanced algorithms must be developed (e.g. vibration control) and will be investigated (WFSs fusion, predictive control, non-linear WF reconstruction for unmodulated WFS and zWFS, etc).

%

\section{END-TO-END PERFORMANCE}
\label{sec:perf}

\subsection{Description of the simulation}

\begin{figure}[t]
    \centering
    \includegraphics[width=\textwidth, trim={0cm 2.cm 62cm 0}]{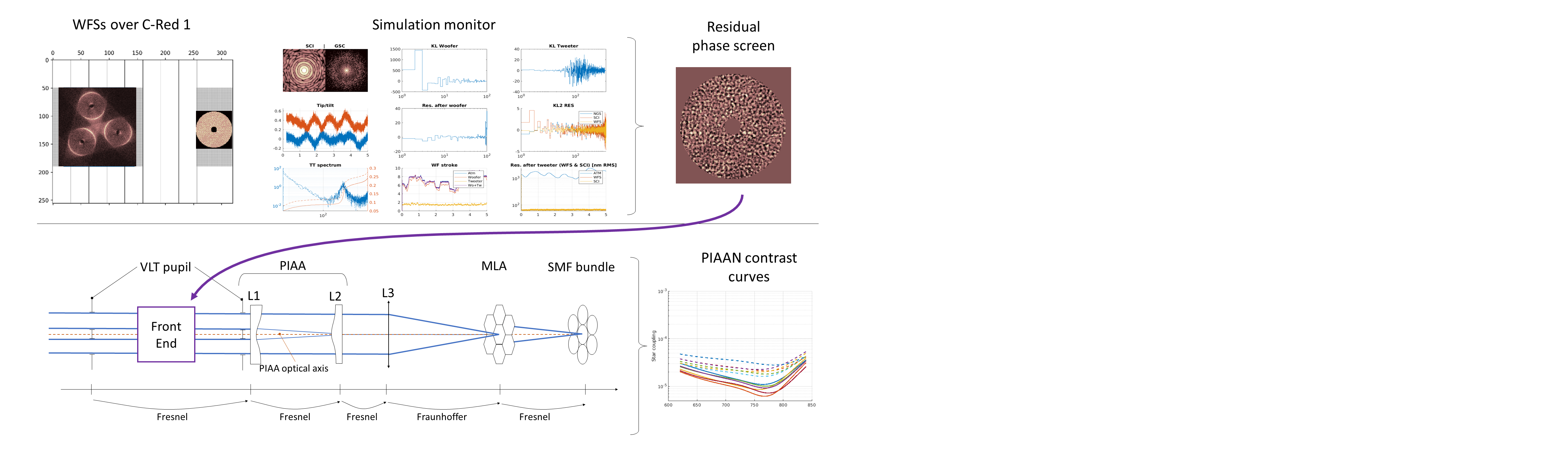}
    \caption{XAO+PIAAN simulation scheme.}
    \label{fig:sim}
\end{figure}

Since last SPIE, we developed a full end-to-end simulation tool. The XAO simulation is based on the OOMAO Matlab library\cite{conan_2014a}. At the end of the XAO simulations, the residual phase screens are injected into the PIAA-Nuller simulations\cite{blind_2022a}, built on top of the Hcipy Python library\cite{por_2018a}, to get coupling over the 7 fibers in the simulated conditions (Fig.~\ref{fig:sim}).

Tab.~\ref{tab:sim_parameters} presents the parameters of the simulations.
Due to some limitation of OOMAO, we set the DM actuator count to 41x41, while oversampling the pupil with both WFSs to 60x60 subapertures. This is potentially the final value for the pyWFS/R-WFS. Thanks to that oversampling, the unmodulated pyWFS can control 1200 modes of the HODM. A higher density DM will improve Strehl, as well as reduce modal confusion. For simplicity we use the same value for the zWFS. We may try to go below 30x30 to save pixels on the C-Red 1, although we have to see if optically speaking we can make both pupils with such different size. Note also that we use the modified version with $\pi/8$ shift, with decreased sensitivity. Both sensors also work at 0-NCPA offset. This is an aspect to be studied, but we will do our best to limit NCPA between WFSs. Finally, we also integrated a Gain Scheduling Camera \cite{chambouleyron_2021a} in the simulation, but don't really use it: an initial correction using the convolutional model with a Fourier simulation is enough. As a confirmation, trying slightly higher integrator gains, in case of optical gain mis-estimation, tends to degrade performance.

For now, we consider the HODM to have an infinite bandwidth. At 4kHz, real DM dynamics could play some role on performance (Sect.~\ref{sec:dm_dynamics}). The dynamics of the ALPAO LODM is modeled with a linear rise time in 1ms (i.e. 2 to 4 frames depending on the simulation sampling). From our experience with the NIRPS DM, and multi-stepping settling, this seems a reasonable approximation. Finally, the HODM is offloaded to the LODM at 500 Hz, which takes into account the LODM rise time to not create oscillations.

The atmosphere is simulated with a 9 layer model from Paranal, to get accurate chromatic pupil shift. Petal modes of different shape can be injected as well (piston-tip-tilt segments, local exponential cooling, etc.): we control them with at most a specific set of 3 piston segments, but no more. Vibrations can also be injected (in particular using measured SPHERE spectra on UT3) but we do not consider them yet, as a controller needs to be implemented \cite{dinis_2024a}.

\begin{table}[t]
    \centering
    \hfill

    \begin{minipage}[t]{0.5\textwidth}
        \caption{Simulations parameters for Prox Cen. Flux includes optical transmission factors (atmosphere, telescope, Front-End, QE).}
        \vspace{0.3cm}
        \begin{tabular}{ll|c|c}
         \hline
         \hline
         && B-WFS & R-WFS \\
         \hline
         WFS        && zWFS  & 3-sided pyWFS \\
         Modulation && $\pi/8$ shift & 0 $\lambda/D$ \\
         $N_{subap}$ && 60x60 & 60x60 \\
         Bandwidth & [nm] & 850-1000 & 1500-1700\\
         $\lambda_{eff}$& [nm] & 925 & 1600 \\
         Prox Cen flux & [ph/ms]  & $1.3\cdot 10^5$ & $4.1\cdot10^5$ \\
         Excess Noise &       & 1.3 & 1.3 \\
         Duty cycle          &       & 0.8 & 0.8 \\
         RON & [$e^-$ RMS] & 0.3 & 0.3 \\
         KL modes & & 1-100 & 101-1200+ \\
         \hline
        Frequency & [kHz] & \multicolumn{2}{c}{2.0 or  4.0} \\
        Delay \& gain & [frame, -] & \multicolumn{2}{c}{[1, 1.0] or [2, 0.4]} \\
         \hline
        \end{tabular}\\
        \vspace{0.3cm}
        \label{tab:sim_parameters}
    \end{minipage}
\hspace{0.05\textwidth} 
    \begin{minipage}[t]{0.35\textwidth}
        \caption{Considered atmospheric conditions for Paranal.}
        \vspace{0.75cm}
        \begin{tabular}{c|cc}
    \hline
    \hline
            Percentile & Seeing at & Wind speed \\
          \%   & zenith ["] & [m/s] \\
        \hline
          10  & 0.52 & 5.2 \\
          25  & 0.62 & 6.3 \\
          50  & 0.76 & 9.5 \\
          75  & 0.96 & 11.7 \\
          90  & 1.26 & 15.0 \\
        \hline
            \end{tabular}
        \label{tab:seeing}
    \end{minipage}
    \hfill

\end{table}

\begin{figure}[t]
    \centering
    \hfill
    \includegraphics[width=0.4\textwidth, trim={18cm 1cm 3cm 0}, clip]{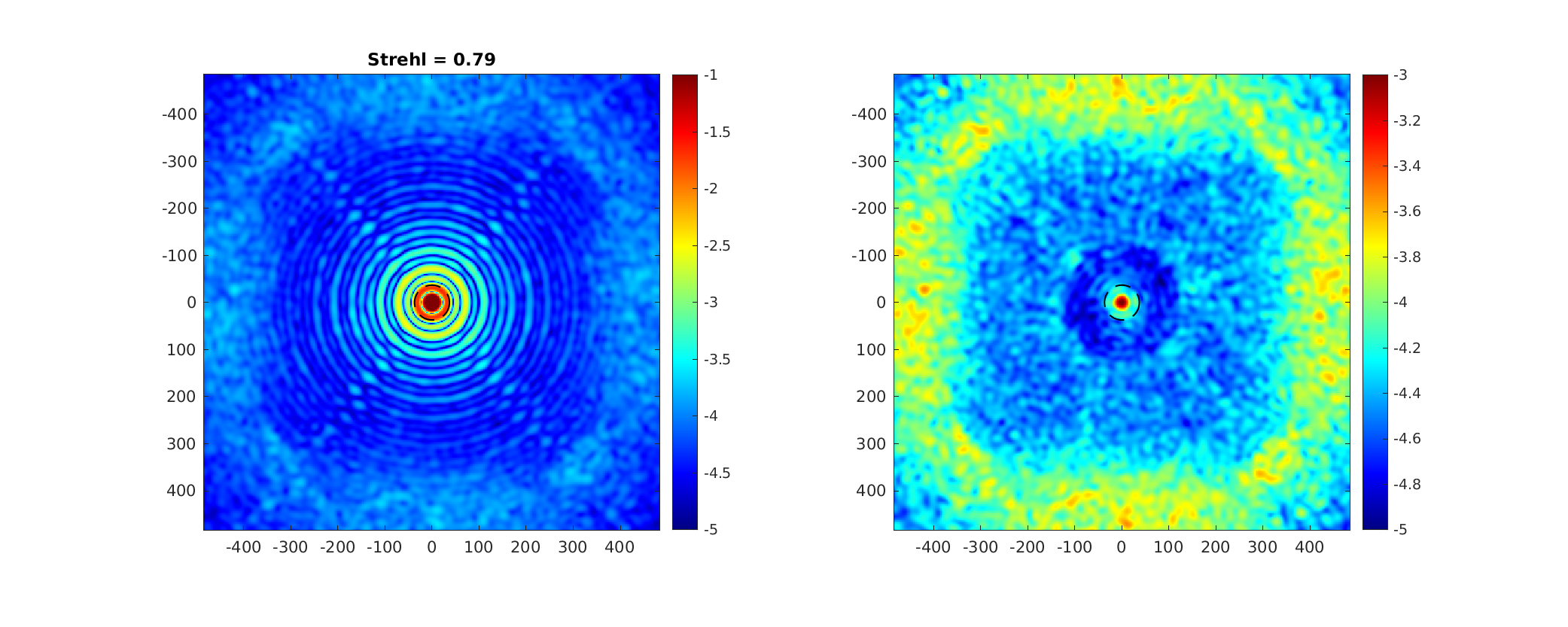}
    \hfill
    \includegraphics[width=0.47\textwidth]{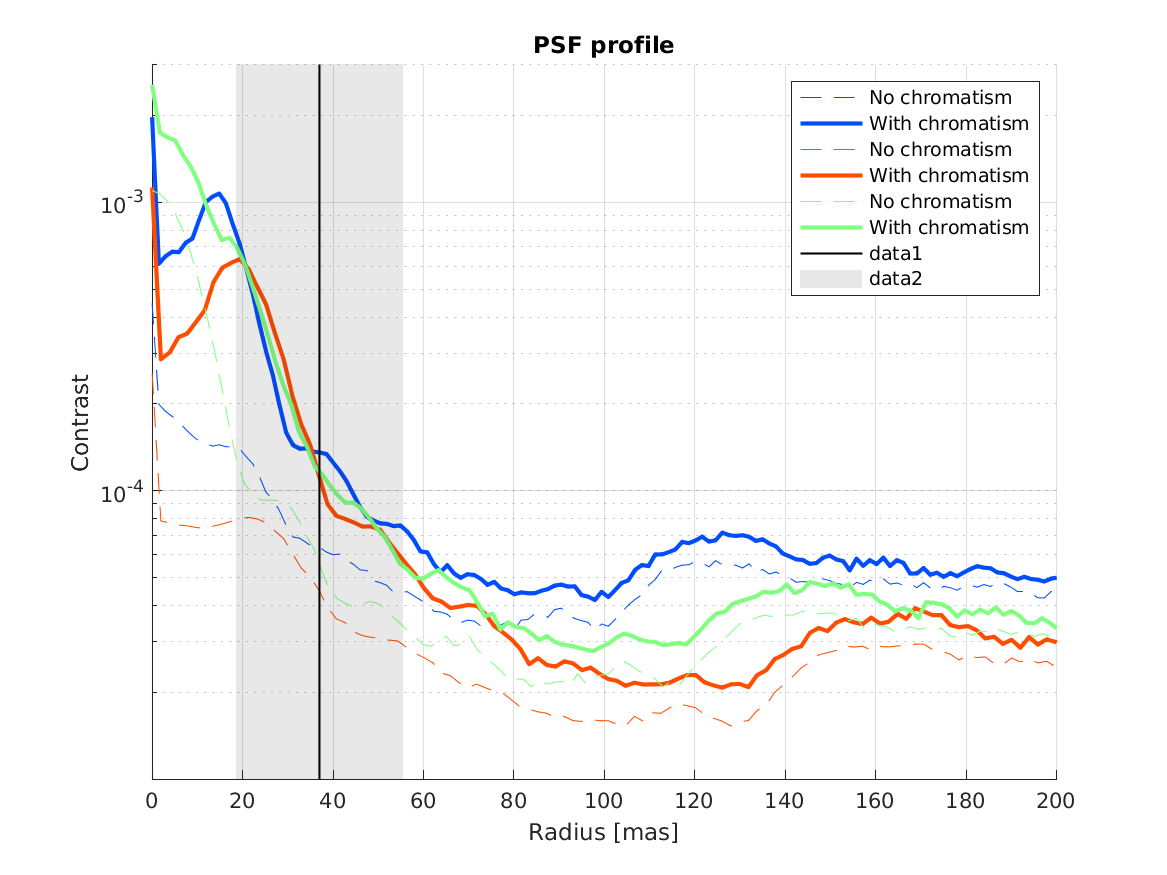}
    \hfill
    \caption{Output of one simulation for a 4kHz, 1-delay XAO, under median seeing conditions and Prox Cen maximum elevation. No fragmentation is injected. Imaging contrast in science band. \textit{Left}: coronagraphic PSF at 730nm. \textit{Right}: radially integrated profiles at 620, 730nm \& 850nm (blue, green, red). Solid line: all included; dashed: without chromatic pupil shift (i.e. considering residual seen by the B-WFS).}
    \label{fig:final_contrast}
\end{figure}

\subsection{RISTRETTO performance}

Fig.~\ref{fig:final_contrast} presents typical result of an OOMAO end-to-end simulation in the case of Prox Cen in median conditions, where we clearly see the zWFS correction zone in the central 100mas. Thanks to the dynamic range in H-band, the unmodulated WFS closes the loop without difficulties even in bad seeing of 1.25" and airmass up to 2. The end-to-end simulations match closely the Fourier model, especially at low spatial frequency. 

We run a grid of simulations for 5 typical atmospheric conditions in Paranal (Tab.~\ref{tab:seeing}), for airmass of 1.00, 1.25 (i.e. the maximum Prox Cen elevation) and 2.00. We test those conditions for 2.0 and 4.0kHz loop speed, with 1 or 2 frame delay, and gains of 0.4 and 1.0 respectively. We end up with 60 simulations per batch, each with a duration of 5s. The output of each simulation is the coupling of the star $\rhos$ into the 6 external fibers, as well as the coupling of a planet $\rhop$ at 18mas centered in one of the 6 external fibers. Coupling is computed for 5 wavelengths between 620 and 840nm (we do not require more, since the residual phase screens quickly smooth down the PIAAN contrast curves). For now, we always use the same seed to generate the phase screens.

\begin{figure}[t]
    \centering
    \includegraphics[width=0.45\textwidth, trim={0 0 0 0}]{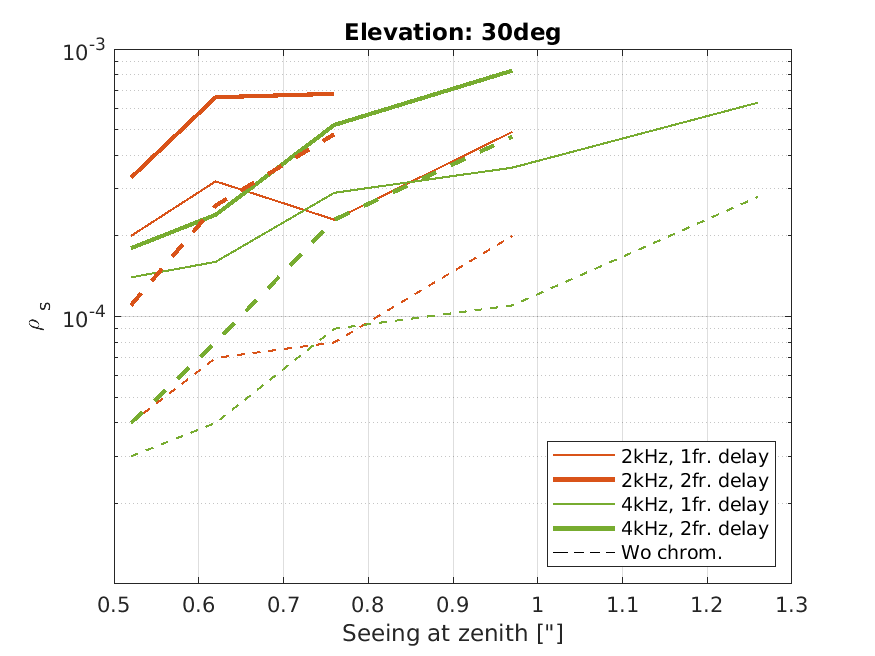}
    \includegraphics[width=0.45\textwidth]{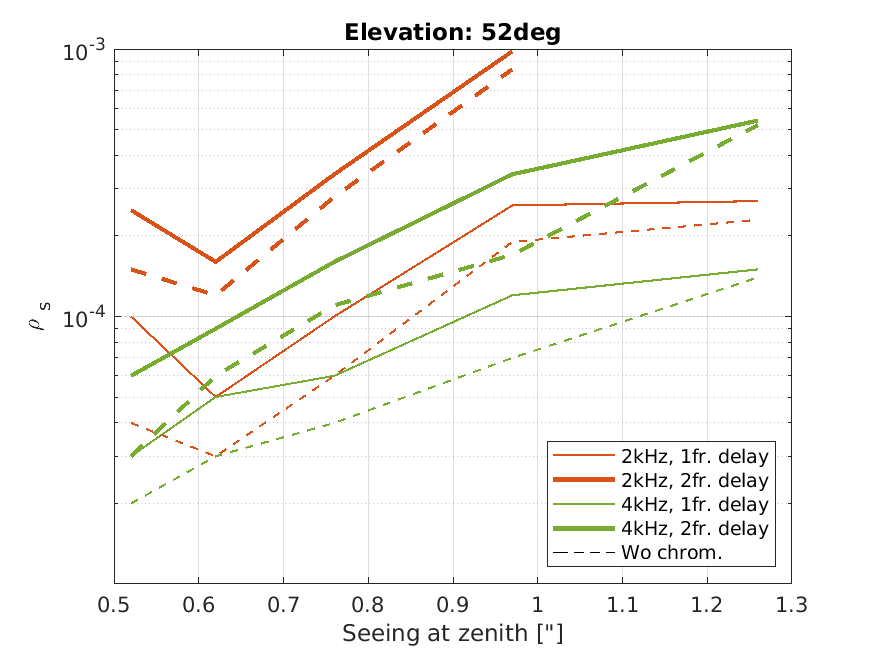}
    \caption{Stellar coupling $\rhos$ in the 6 external fibers at the end of the end-to-end simulation presented in Fig.~\ref{fig:sim}. Values are averaged over 5 wavelengths and the 6 fibres. We considered 2 loop speed (green and orange) and 2 delay values (thin \& thick lines) under 5 seeing conditions. Solid lines are all inclusive, while dotted lines ignore the chromatic pupil shift. Missing points at high seeing correspond to failed simulations.}
    \label{fig:final_perf}
\end{figure}

Fig.~\ref{fig:final_perf} summarizes results of those runs for elevations of 52$^\circ$ and 30$^\circ$. We only show here $\rhos$, which is the most strongly varying value among all simulations. $\rhop$ is proportional to Strehl: it varies little between the 4 XAO configurations, and decreases with increasing seeing and airmass. Each point in those curves is the average of $\rhos$ over the 6 external fibers and the 5 wavelengths. This hides fiber-to-fiber differences due to the wind directions in some situations.

For sake of simplicity, let's consider here to reach our specification if $\rhos \le 1\cdot10^{-4}$. In median conditions, specification is not reached by our baseline XAO running at 2kHz. Null only reaches $\rhos \sim2\cdot10^{-4}$, instead of $\rhos \sim 0.5\cdot10^{-4}$ expected from Fourier simulations. Accounting for $\rhop$, the SNR is 40\% lower than the specification (i.e. integration time 2.5 times longer). From the coronagraphic images, or the PSD of residual phase screens, lag appears as the dominant term. It only disappears considering lower lag simulations, i.e. 2kHz with predictive controller and/or faster 4kHz system appear necessary. Note also that in simulation performed without photon noise, the central area is well corrected by the unmodulated pyramid, pointing to photon noise limit on the pyramid sensor.

In the best conditions (10\% of the time), those simulations suggest we could achieve a potential cut in integration time by a factor of 4-5 compared to the requirement, gain limited by chromaticity.

Also, the impact of chromatic pupil shift even at max Prox Cen elevation is noticeable, reducing contrast by a factor $\sim$2 in best situations. Contrary to the results of the Fourier simulation, B-WFS appears already too far from the science band to manage the chromatic pupil shift at airmass of 2. This could nevertheless be caused by a bad atmospheric draw, chromatic pupil shift applied in the direction of the dominant wind, or an inaccurate modeling of the effect. Work is still on-going on that aspect\cite{motte_2024a}.

\begin{figure}[b]
    \vspace{0.3cm}
    \centering
    \includegraphics[width=0.9\textwidth, trim={0 0 16cm 0}, clip]{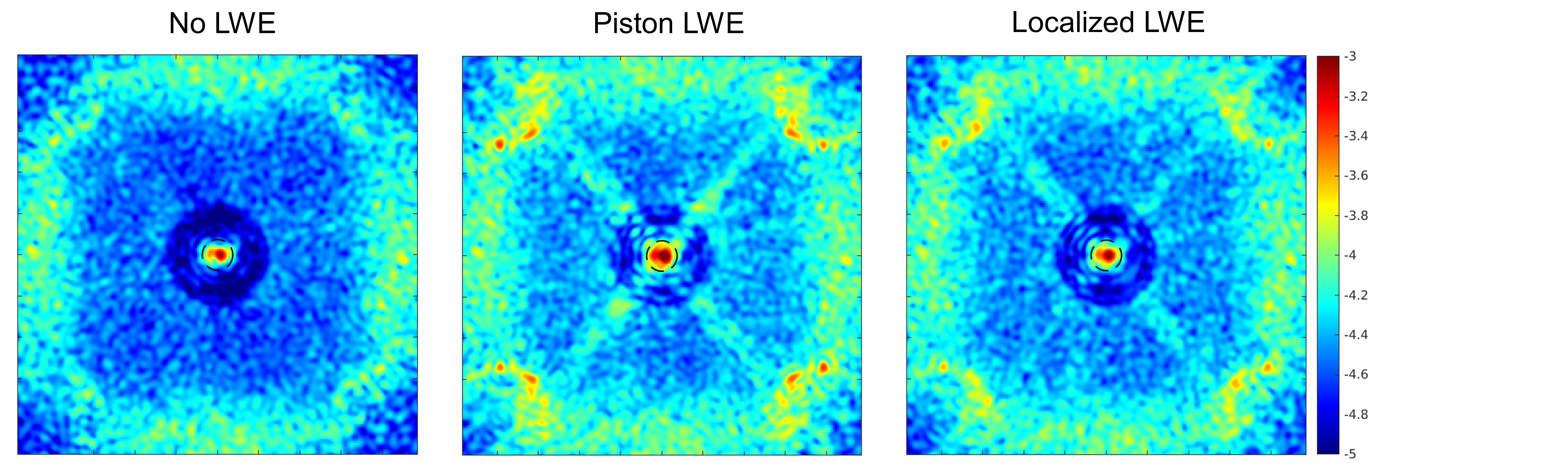}
    \caption{Coronagraphic PSFs at 750nm in presence with or without low wind effect (\textit{piston lwe} corresponds to \textit{local lwe} with infinite cooling zone).  Atmospheric conditions set to 10-percentile best seeing conditions, at 52$^\circ$ elevation. }
    \label{fig:petal_control}
\end{figure}

\subsection{Some notes on zWFS stability and petal control}

The zWFS is more and more considered for 2nd stage XAO system. However, it still proves to be a difficult, sensor even in simulations due to its small dynamic range.

Over the grid of 60 simulations, under pure Kolmogorov perturbation, the $\pi/8$-zWFS failed (loop diverging) in 5 of them, when lag and seeing are worst (seeing $\ge 1.25"$, f=2kHz  \& delay=2 frame). Such situation is probably not critical for RISTRETTO, since they are of too low SNR, but they point us towards the limiting use conditions of the zWFS. Reducing the number of modes controlled by the zWFS to 50 reduced the number of failed simulations to 3, without impacting significantly performance, due to the very narrow working angle. Also, running the same simulation with the standard $\pi/2$-zWFS significantly increases failure rate to 16 out of the 60 simulations, including some of the lowest latency, 4kHz \& 1-delay simulations. Performance being limited by the lag term, the higher phase shift does not improve final performance on Prox Cen, and gets worse as we approach the unstable regime.

Finally, as mentioned earlier, zWFS can control the petal modes to required level (Fig.~\ref{fig:petal_control}). To test it, we inject low wind disturbance at high frequencies (between 1 and 10Hz) with amplitudes up to 1$\mu m$, as reported in different papers. We inject up to 3 such modes at the same time (either pure piston, or local cooling), with maximum amplitude from the start of the simulation (so the R-WFS has to manage them first). If we inject as \textit{petal}, the best DM fit (and not a sharp piston-segment), close-loop performance is on par to simulations without petal at all. If we introduce now a real discontinuity (a piston mode or a local cooling effect), the impact is noticeable but limited on the fiber contrast. We observe a higher failure rate than without the effect, but only for the worst seeing $\ge 1.26$. Considering the lowest wind speed case, where the effect has the most chances to appear, $\rhos$ increases from $0.3\times10^{-4}$ to $0.6\times10^{-4}$ in presence of low wind effect. This increase appears mostly due to the DM fit residual to the discontinuity, and not to bad sensing. Removing the atmosphere, this performance is limited to $\sim 0.4\times10^{-4}$. Given the occurrence rate of the effect, and the fact that the loop is stable, this degradation is acceptable. We have several ideas to manage it on the PIAAN side or at the WFS level if it proves problematic with more advanced simulations.

We also simulated an unmodulated pyWFS as B-WFS: while it behaves well under pure Kolmogorov, it shows an erratic behavior in presence of petal modes. This erratic behavior is observed even under good seeing and in absence of photon noise: we therefore believe it is due to the before mentioned cross-talks between discontinuity signal on each side of the spider. Even in absence of atmosphere, while stable, it performs poorly and does not reach the desired contrast.

\section{Front-End preliminary design}
\label{sec:fe}

\begin{figure}
    \centering
    \includegraphics[width=0.95\textwidth]{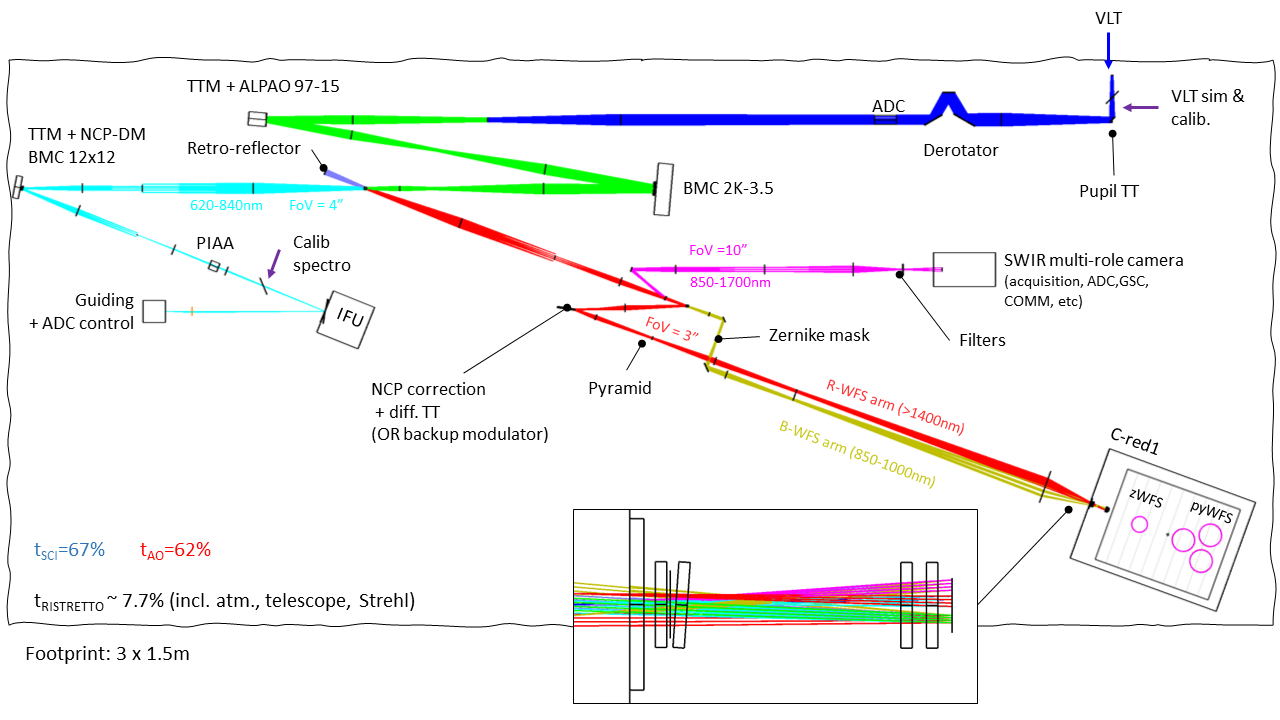}
    \caption{Preliminary Front-End paraxial design.}
    \label{fig:fe_design}
\end{figure}

We recently established a preliminary (paraxial) optical design of the Front-End (Fig.~\ref{fig:fe_design}), to estimate occupation at the VLT Nasmyth platform, shared with the spectrograph and all the cabinets. It was also the occasion to study the possibility to fit a zWFS and a pyramid WFS together on the same Cred-1 detector. Our baseline is to use refractive optics, but no definitive choice can be made for now.

\section{CONCLUSION}
\label{sec:ccl}

Our most recent end-to-end simulations demonstrate the performance of RISTRETTO to the required level, although for now we need a better system than anticipated. Those results suggest to seriously develop more advanced linear WF estimators and advanced, predictive control laws for the zWFS and low order modes. 

RISTRETTO is a challenge forcing us to explore and demonstrate new concepts up to an operational point. Characterisation of some key hardware has already started (C-Red 1 and Boston DM). The simulations have reached an excellent complexity and maturity level, and it is time to also confront us to real life. We plan on starting prototyping activities of unmodulated pyramid \& zWFS in 2024-2025.  Activities around fundamental sub-systems such as the ADC and the RTC should also start soon.

Finally, note that the PIAAN prototyping is already progressing at good pace, with objective to demonstrate its performance as well as low-order control by the end of 2024.

\section*{Acknowledgement}
This work has been carried out within the framework of the National Centre of Competence in Research PlanetS supported by the Swiss National Science Foundation under grants 51NF40\_182901 and 51NF40\_205606. The RISTRETTO project was partially funded through SNSF FLARE programme for large infrastructures under grants 20FL21\_173604 and 20FL20\_186177. The authors acknowledge the financial support of the SNSF.

\bibliography{references} 
\bibliographystyle{spiebib} 

\end{document}